\documentclass[superscriptaddress,secnumarabic,amssymb,amsmath,12pt,nobibnotes,aps,prd,showkeys,nofootinbib]{revtex4-1}
\usepackage{amsmath}

\usepackage{graphicx}
\usepackage{epsf}
\usepackage{bm}
\usepackage{amsmath}
\usepackage{amsfonts}
\usepackage{amssymb}
\usepackage{epstopdf}
\usepackage{natbib}
\usepackage{color}

\begin{document}
\tolerance=5000
\noindent
%--------------------------------------------------------------------
\title{Scaling solutions in quintessential inflation}

\author{Jaume Haro}
\email{jaime.haro@upc.edu}
\affiliation{Departament de Matem\`{a}tiques, Universitat Polit\`{e}cnica de Catalunya, Diagonal 647, 08028 Barcelona, Spain}

\author{Jaume Amor\'{o}s}
\email{jaume.amoros@upc.edu}
\affiliation{Departament de Matem\`{a}tiques, Universitat Polit\`{e}cnica de Catalunya, Diagonal 647, 08028 Barcelona, Spain}

\author{Supriya Pan}
\email{supriya.maths@presiuniv.ac.in}
\affiliation{Department of Mathematics, Presidency University, 86/1 College Street, Kolkata 700073, India.}

\begin{abstract}
In quintessence scalar field theories, the presence of scaling solutions are important during the radiation and matter epoch due to having their attractor character. Usually, it is assumed that the initial conditions of the quintessence field are in the basin of attraction of the scaling solutions. However, in order to reproduce the current cosmic acceleration, at late times, 
a mechanism to exit this behavior  
is needed. In the present work we show that the quintessential inflation models could be an excellent candidate to exhibit the above behavior. 
However, the crucial point of  quintessential inflation is that the initial conditions has to be taken during the inflation,  and  at the beginning of the radiation era,  the scalar field does not belong to the basin of attraction of the scaling solution. This  means that, in the case where quintessence is depicted via exponential potentials,  only a single exponential in the tail of the quintessential inflation potential is enough to depict the evolution of our universe. 

\end{abstract}

\vspace{0.5cm}

%-------------------------------------------------------------------------
\maketitle
%------------------------------------------------------------------------
\thispagestyle{empty}
%\tableofcontents

\section{Introduction}

In standard inflation \cite{guth, linde},
after the decay of the inflaton field, and subsequent beginning of 
the radiation dominated epoch, considering a quintessence scalar field \cite{Caldwell:1997ii} whose potential is exponential, $V(\varphi)=V_0e^{-\gamma\varphi/M_{pl}}$, with
$\gamma>2$, there exists a solution whose energy density scales as the one of radiation \cite{copeland}. Such solution is termed as the scaling solution. In fact, for a more general class of exponential potentials of the form, 
 $V(\varphi)=V_0e^{-\gamma\varphi^n/M_{pl}^n}$, there exists an approximate scaling solution \cite{Geng:2017mic}.

Due to the attractor behavior of this solution, at the beginning of the radiation era, the initial conditions of the quintessence field belonging to the basin of attraction of this scaling solution could be used to deal with the so called {\it coincidence of scales}. Obviously, in order to reproduce the late time acceleration of the universe, the quintessence field has to leave the scaling behavior, which could be done in several ways. Taking into account that, for $0<\gamma<\sqrt{2}$, during the matter domination era there exists a tracker solution \cite{Steinhardt:1999nw, rp, copeland} leading to an accelerating late time universe,
 one could consider a double exponential potential
 $V(\varphi)=V_re^{-\gamma_r\varphi/M_{pl}}+ V_me^{-\gamma_m\varphi/M_{pl}}$, with $\gamma_r>2$ and $0<\gamma_m<\sqrt{2}$. The first exponential function of this double exponential potential dominates during the radiation dominated era and the second exponential potential dominates during the matter dominated era \cite{barreiro}, and gives an effective $\gamma(\varphi)$ varying smoothly between two desired values. Alternatively, one could introduce a non-minimal coupling between the quintessence field and massive neutrinos, whose effect is to modify the potential in the matter domination era \cite{hossain2,Geng:2017mic}.

On the contrary, in quintessential inflation \cite{pv,dimopoulos1,hossain1, hap1,deHaro:2016hsh,deHaro:2016ftq,hap,deHaro:2017nui,AresteSalo:2017lkv,Haro:2015ljc,haro19,hap19}, there is only one scalar field  driving the evolution of the universe by depicting both the early- and  late- acceleration of the universe. Due to the attractor behavior of inflation the initial condition of the scalar field has to be taken belonging to the basin of attraction of the slow roll solution. Then, using a quintessential inflation model based in an exponential SUSY inflation model \cite{stewart} matched with an exponential potential with  $\gamma>2$, which acts as a quintessential tail (although our reasoning does not depend on the inflationary piece of the potential), 
we show that at the beginning of the radiation era the scalar field is not in the basin of attraction of the scaling solution. In fact, the value of 
the ratio of the energy density of the scalar field to the critical energy, namely
$\Omega_{\varphi}$, is very low and satisfies the bounds imposed at the Big Bang Nucleosynthesis (BBN) and at the recombination \cite{Geng:2017mic}. 
As a consequence, in quintessential inflation a mechanism to exit the scaling behavior is not needed. The only thing that is needed is an inflationary potential leading to a spectral index ($n_s$) and the ratio of tensor-to-scalar perturbations ($r$) entering into the two dimensional marginalized joint confidence contour at $2\sigma$ confidence-level (CL) provided by Planck data (for example, 
for the Planck 2018 TT, TE, EE + low E+ lensing + BK14 + BAO likelihood \cite{planck18,planck18a}), matched with a quintessence tail, whose potential, in order to have a kination regime after inflation \cite{Joyce},  should be small compared to the kinetic energy density of the field at the end of inflation. Additionally,  at late times, the scalar field potential has to be dominant with respect to the energy density of the matter, for example for the exponential potential, $V(\varphi)=V_0e^{-\gamma\varphi/M_{pl}}$, with $0<\gamma<\sqrt{2}$ and $V_0\ll M_{pl}^4$.

The paper is organized as follows. In Section \ref{sec-II}, we consider a universe filled with a barotropic fluid and a scalar field with an exponential potential, and we find the scaling and tracker solutions. Section \ref{sec-III} is devoted to the study of a model of quintessential inflation potential whose quintessential tail is an exponential one with $\gamma>2$, showing analytically and numerically that at the beginning of radiation, the scalar field does not belong to the basin of attraction of the scaling solution. In Section \ref{sec-IV}, we propose a viable model based on the matching of an exponential SUSY  inflationary model and an exponential potential with $0<\gamma<\sqrt{2}$. Finally, in Section \ref{sec-summary}, we present the conclusions of our work.

The units used throughout the paper are, $\hbar=c=1$, and we denote  the reduced Planck's mass by 
$M_{pl}\equiv \frac{1}{\sqrt{8\pi G}}\cong 2.44\times 10^{18}$ GeV.

\section{Scaling and tracker  solutions}
\label{sec-II}

Assuming, as usual, the   flat  Friedmann-Lema\^{i}tre-Robertson-Walker (FLRW) geometry of our universe, in this Section we consider an exponential potential, $V(\varphi)=V_0e^{-\gamma \varphi/M_{pl}}$, and  a barotropic fluid with Equation of State (EoS) parameter $w=1/3$, i.e., a radiation fluid,  whose energy density  we denote by $\rho_r$. Then, the dynamical system is given by the equations
\begin{eqnarray}\left\{ \begin{array}{ccc}
\ddot{\varphi}+3H\dot{\varphi}+V_{\varphi}&=& 0\\
\dot{\rho}_r+4H\rho_r &=& 0,
\end{array}\right.
\end{eqnarray}
where $H=\frac{1}{\sqrt{3}M_{pl}}\sqrt{\rho_{\varphi}+\rho_r}$, is the Hubble rate of the FLRW universe with $\rho_{\varphi}=\frac{\dot{\varphi}^2}{2}+V(\varphi)$, being the energy density of the scalar field.

The scaling solution, is a solution with the property that the energy density of the scalar field scales as the one of radiation, meaning that the Hubble parameter evolves as, $H(t)=\frac{1}{2t}$.  At this point, we look for a solution of the form
\begin{eqnarray}
\varphi_{sc}(t)=\frac{M_{pl}}{\gamma}\ln\left( \frac{t^2}{\bar{t}^2} \right), \qquad \rho_r(t)=\frac{C^2}{t^2},
\end{eqnarray}
where $C$ and $\bar{t}$ are constants.
The equation, $\dot{\rho}_r+4H\rho_r = 0$, is satisfied for any value of $C^2$ and the equation $\ddot{\varphi}+3H\dot{\varphi}+V_{\varphi}= 0$ requires that
$\bar{t}^2=\frac{M_{pl}^2}{\gamma^2V_0}$.

On the other hand,  equating $H(t)=\frac{1}{2t}$ with $H=\frac{1}{\sqrt{3}M_{pl}}\sqrt{\rho_{\varphi}+\rho_r}$, that yields the following relation
\begin{eqnarray}
C^2=\frac{3}{4}\left(1-\frac{4}{\gamma^2} \right)M_{pl}^2,
\end{eqnarray}
shows that the scaling solution only exists for $\gamma>2$. Then, the corresponding energy densities evolve as follows:
\begin{eqnarray}
\rho_{\varphi_{sc}}(t)=\frac{3M_{pl}^2}{\gamma^2t^2} \qquad \rho_r(t)= \frac{3}{4}\left(1-\frac{4}{\gamma^2} \right)\frac{M_{pl}^2}{t^2},
\end{eqnarray}
and thus, one could derive the corresponding density parameters:
\begin{eqnarray}
\Omega_{\varphi_{sc}}=\frac{4}{\gamma^2},\qquad \Omega_r=1-\frac{4}{\gamma^2}.\end{eqnarray}

An alternative approach to get this solution goes as follows
(see \cite{copeland}): we introduce the dimensionless variables 
\begin{eqnarray}
\tilde{x}\equiv \frac{\dot{\varphi}}{\sqrt{6}M_{pl}H} \quad \mbox{and} \quad \tilde{y}\equiv \frac{\sqrt{V}}{\sqrt{3}M_{pl}H},
\end{eqnarray}
which enable us to write down the following autonomous dynamical system
 \begin{eqnarray}\label{dynamical}
 \left\{\begin{array}{ccc}
 \tilde{x}'&=& -3\tilde{x}+\sqrt{\frac{3}{2}}\gamma \tilde{y}^2+\frac{3}{2}\tilde{x}\left[(1-w)\tilde{x}^2+(1+w)(1-\tilde{y}^2)\right]\\
  \tilde{y}'&=& -\sqrt{\frac{3}{2}}\gamma\tilde{x}\tilde{y}+\frac{3}{2}\tilde{y}\left[(1-w)\tilde{x}^2+(1+w)(1-\tilde{y}^2)\right]\ \end{array},
 \right.
 \end{eqnarray}
 where $w$ is the Equation of State (EoS) parameter, which in our case is equal to $1/3$, 
together with the constraint
\begin{eqnarray}
\tilde{x}^2+\tilde{y}^2+\Omega_r=1.
\end{eqnarray}

One can see that, for $w=1/3$,  the dynamical system (\ref{dynamical}) has the following attractor solution $\tilde{x}=\left(\sqrt{\frac{2}{3}}\; \right)\frac{2}{\gamma},\quad \tilde{y}=\frac{2}{\sqrt{3}}\frac{1}{\gamma}$, whose energy density scales as radiation and $\Omega_{\varphi}=\frac{4}{\gamma^2}$.

So, since during radiation $H=\frac{1}{2t}$, thus, from $\tilde{x}=\left(\sqrt{\frac{2}{3}}\; \right)\frac{2}{\gamma}$, we have 
\begin{eqnarray}
\dot{\varphi}_{sc}=\frac{2}{\gamma t}M_{pl} \Longrightarrow \varphi_{sc}(t)=\frac{M_{pl}}{\gamma}\ln\left( \frac{t^2}{\bar{t}^2}\right).
\end{eqnarray}

Finally, in order to obtain $\bar{t}$, we use the equation $\tilde{y}=\frac{2}{\sqrt{3}}\frac{1}{\gamma}$, which leads to $\bar{t}^2=\frac{M_{pl}^2}{\gamma^2V_0}$, and thus, 
we recover the scaling solution 
\begin{eqnarray}
\varphi_{sc}(t)=\frac{M_{pl}}{\gamma}\ln\left( \frac{\gamma^2V_0t^2}{M_{pl}^2}  \right).
\end{eqnarray}

\subsection{The tracker solution}
\label{sec-III}

A tracker solution \cite{Steinhardt:1999nw, UrenaLopez:2000aj}, is an attractor solution of the field equation which describes the
dark energy domination at late time. In fact, for an exponential potential, it is the solution of $\ddot{\varphi}+3H\dot{\varphi}+V_{\varphi}=0$, when the universe is filled by the scalar field only.
Once again, we look for a solution of the type
\begin{eqnarray}
\varphi_{tr}(t)=\frac{M_{pl}}{\gamma}\ln\left( \frac{t^2}{\bar{t}^2} \right).
\end{eqnarray}

Inserting this equation into $\ddot{\varphi}+3H\dot{\varphi}+V_{\varphi}=0$, one gets
\begin{eqnarray}
\bar{t}^2=\frac{2M_{pl}^2}{V_0\gamma^4}(6-\gamma^2),
\end{eqnarray}
which means that, $0<\gamma <\sqrt{6}$, and thus, the tracker solution is given by 
\begin{eqnarray}\label{tracker}
\varphi_{tr}=\frac{M_{pl}}{\gamma}\ln\left(\frac{V_0\gamma^4}{2(6-\gamma^2)M_{pl}^2}t^2   \right).
\end{eqnarray}

Moreover, for this equation, it is not difficult to show that the corresponding effective EoS parameter is given by
\begin{eqnarray}
w_{eff}=\frac{\gamma^2}{3}-1,
\end{eqnarray}
which demonstrates that to realize the late time acceleration of the universe, we need to restrict $\gamma$ as 
$0<\gamma<\sqrt{2}$.

Note also that the system (\ref{dynamical}) when $w=0$ (in the matter domination era), has the fixed point $(\tilde{x}, \tilde{y})= \left(\gamma/\sqrt{6}, \sqrt{1- \gamma^2/6} \right)$, which of course, corresponds to the solution  (\ref{tracker}).

\section{A quintessential inflation model}
\label{sec-III}

In this Section 
we consider an Exponential SUSY Inflation-type potential  (see \cite{martin} and references therein) matched with an exponential potential as follows:
\begin{eqnarray}\label{SUSY}
V(\varphi)=\left\{\begin{array}{ccc}
\lambda M_{pl}^4\left( 1-e^{\varphi/M_{pl}} + \left(\frac{M}{M_{pl}}\right)^4\right) & \mbox{for} & \varphi\leq 0\\
\lambda{M^4}e^{-\gamma\varphi/M_{pl}}&\mbox{for} & \varphi\geq 0,\end{array}
\right.
\end{eqnarray}

For this model the spectral index and the ratio of tensor-to-scalar perturbations, as a function of the number of e-folds $N$, are given by 
\begin{eqnarray}
n_s\cong 1-\frac{2}{N},\qquad r\cong \frac{8}{N^2},
\end{eqnarray}
which shows that for a number of $e$-folds  greater than $60$, the ratio of tensor to scalar perturbations is less than $0.003$.  Thus, the joint contour of ($n_s$, $r$) enters perfectly into the $1\sigma$ CL for the Planck2018 TT, TE, EE + low E+ lensing + BK14 + BAO likelihood.

Now, in order to calculate the value of the parameter $\lambda$, we use that for this model the power spectrum of scalar 
perturbations is given by
\begin{eqnarray*}
{\mathcal P}_{\zeta}\cong \frac{\lambda}{12\pi^2}e^{-2\varphi_*/M_{pl}}\sim 2\times 10^{-9}.
\end{eqnarray*}

Then, using $e^{\varphi_*/M_{pl}}\cong \frac{1-n_s}{2}$, and taking  the central value of the spectral index  $n_s\cong 0.96$ \cite{planck18}, one gets
\begin{eqnarray}
\lambda\cong 6\pi^2(1-n_s)^2\times 10^{-9}\cong 6\times 10^{-11}.
\end{eqnarray}

On the other hand, 
inflation ends at $\varphi_{end}=-\ln\left(1+\frac{1}{\sqrt{2}}  \right)M_{pl}$. Thus, the energy density at the end of inflation is
\begin{eqnarray}\rho_{\varphi, end}\cong V(\varphi_{en})=\frac{\lambda}{1+\sqrt{2}}M_{pl}^4\cong 2.5\times 10^{-11} M_{pl}^4.
\end{eqnarray}

Therefore,
assuming there is no drop of energy between the end of inflation and the beginning of kination, one has
\begin{eqnarray}
H_{kin}\cong 2.9\times 10^{-6} M_{pl} \qquad \dot{\varphi}_{kin}\cong  7\times 10^{-6} M_{pl}^2.
\end{eqnarray}

Next,  we want to calculate the value of the scalar field and its derivative at the reheating time. To simplify our calculation we assume that the reheating is due to instant preheating, although  our reasoning will be the same  when the reheating is via the gravitational production of superheavy particles  \cite{haro18,hashiba},
which means that the created particles have to decay in lighter ones before the end of the kination phase \cite{fkl0, fkl}.

Analytical calculations could be done disregarding the potential during kination. Then,  since during kination one has, $a\propto t^{1/3}\Longrightarrow H=\frac{1}{3t}$, and consequently, using the Friedmann equation, the evolution of the scalar field in this regime will be governed by
\begin{eqnarray}
\frac{\dot{\varphi}^2}{2}=\frac{M_{pl}^2}{3t^2}\Longrightarrow \dot{\varphi}=\sqrt{\frac{2}{3}}\frac{M_{pl}}{t}\Longrightarrow 
\varphi(t)=\sqrt{\frac{2}{3}}M_{pl}\ln \left( \frac{t}{t_{kin}} \right).\end{eqnarray}

At the reheating time, one has 
\begin{eqnarray}
\varphi_{rh}=\sqrt{\frac{2}{3}}M_{pl}\ln\left( \frac{H_{kin}}{H_{rh}} \right),
\end{eqnarray}
and using that at the reheating time, i.e., when the energy density of the scalar field and the one of the relativistic plasma coincide: $H_{rh}^2=\frac{2\rho_{rh}}{3M_{pl}^2}$, one gets 
\begin{eqnarray}
\varphi_{rh}=\sqrt{\frac{2}{3}}M_{pl}\ln\left( \frac{2.9\times 10^{-6} M_{pl}}{\sqrt{\frac{\pi^2g_{rh}}{45}} \frac{T_{rh}^2}{M_{pl}}}\right),
\qquad \mbox{and}  \qquad  { \dot{\varphi}_{rh}=\sqrt{\frac{2\pi^2g_{rh}}{15}} T_{rh}^2},\end{eqnarray} 
where we have used that, at the reheating time, the energy density and the temperature are related via 
 $\rho_{rh}=\frac{\pi^2}{30}g_{rh}T_{rh}^4$, where the number of degrees of freedom for the Standard Model is $g_{rh}=106.75$  \cite{rg}. In addition, to obtain $\dot{\varphi}_{rh}$, we have used that $\dot{\varphi}=\sqrt{\frac{2}{3}}\frac{M_{pl}}{t_{rh}}$, and since $H_{rh}=\frac{1}{3t_{rh}}$, we get
 $\dot{\varphi}=\sqrt{6}H_{rh}M_{pl}=2\sqrt{\rho_{rh}}=
 \sqrt{\frac{2\pi^2g_{rh}}{15}} T_{rh}^2$.

Since when the reheating is via instant preheating, the  reheating temperature is of the order of $10^9$ GeV \cite{haro19}, and one will have
 \begin{eqnarray}
 \varphi_{rh}\cong 23.6 M_{pl} \qquad {\dot{\varphi}_{rh}\cong 1.45\times 10^{-18} M_{pl}^2}.
 \end{eqnarray}

On the other hand, at the beginning of the radiation era, the value of the scaling solution and its derivative are given by 
\begin{eqnarray}
\varphi_{sc,rh}=\frac{M_{pl}}{\gamma}\ln\left( \frac{\lambda\gamma^2 M^4}{4H_{rh}^2M_{pl}^2}  \right)
\qquad \dot{\varphi}_{sc,rh}=\frac{4}{\gamma}H_{rh}M_{pl},
\end{eqnarray}
which for $T_{rh}\cong 10^9$ GeV, leads to
\begin{eqnarray}
\varphi_{sc,rh}\cong \frac{M_{pl}}{\gamma}\left(9.2n+36+2\ln\gamma  \right)
\qquad \dot{\varphi}_{sc,rh}\cong 3.2 \times 10^{-18}M_{pl}^2,
\end{eqnarray}
where we have written $M=10^n M_{pl}$, $n$ being an integer.

Now, taking into account that the upper limit from BBN suggests that  the scaling solution has to satisfy 
$\Omega_{\varphi_{sc}}(T=1 \mbox{ MeV})< 0.045$ \cite{bean}, hence, in order that the scaling solution satisfies this bound,
and choosing for example $\gamma=10$, we have 
$\varphi_{sc,rh}\cong \frac{M_{pl}}{10}(9.2n+40.6)$. Then, if one wants that the real value of the scalar field was near to the value of the scaling solution at the reheating time, the following relation has to be accomplished
\begin{eqnarray}
9.2n+40.6\cong 236 \Longrightarrow n\cong 21,
\end{eqnarray}
which is completely unacceptable, because this means $M\cong 10^{21} M_{pl}$  and thus, it is impossible to have a kination regime ($M\ll M_{pl}$) at the end of inflation
(In fact, for $\varphi_{rh}\cong 23.6 M_{pl}$, it is impossible to  accomplish the kinetic regime for any value of $\gamma>2$, because one always obtains a positive value of $n$, meaning that  $M\gg M_{pl}$, which forbids a kination regime.

This means that, at the beginning of the reheating the real solution is completely different to the scaling one, and, as we will see numerically,  at the matter-radiation equality,
both solutions continue to be different, which means that the real solution does  not belong to the basin of attraction of the scaling solution.

A key point  to understand this fact is that during all the radiation regime the kinetic energy of the scaling solution is twice its potential energy, however, for the real solution, at the beginning
of reheating: $\frac{\dot{\varphi}_{rh}^2}{2}\gg V(\varphi_{rh})$, and this property, as we will show numerically soon, is conserved during all the radiation dominated phase. The conclusion is that, in quintessential inflation the scaling regime is never reached for the scalar filed.  To show this analytically one has to take into account that, during this period, to obtain the evolution of the scalar field, due to the initial conditions at the beginning of the radiation era, on could continue disregarding the potential, obtaining
 \begin{eqnarray}
 \varphi(t)=\varphi_{rh}+2\dot{\varphi}_{rh}t_{rh}\left( 1-\sqrt{\frac{t_{rh}}{t}} \right).
 \end{eqnarray}
 
 To calculate the value of the field and its derivative at the matter-radiation equality, i.e., $\varphi_{eq}$ and $\dot{\varphi}_{eq}$, we consider the central values  obtained in \cite{planck} (see the second column in Table $4$) of the redshift at the matter-radiation equality  $z_{eq}=3365$, 
the present value of the ratio of the matter energy density to the critical one $\Omega_{m,0}=0.308$, and $H_0=67.81\; \mbox{Km/sec/Mpc}=5.94\times 10^{-61} M_{pl}$.
Then, the present value of the matter energy density is, $\rho_{m,0}=3H_0^2M_{pl}^2\Omega_{m,0}=3.26\times 10^{-121} M_{pl}^4$, and at the matter-radiation equality we have $\rho_{eq}=2\rho_{m,0}(1+z_{eq})^3=2.48\times 10^{-110} M_{pl}^4=8.8\times 10^{-1} \mbox{eV}^4$. 
Now,  using the relation at the matter-radiation equality $\rho_{eq}=\frac{\pi^2}{15}g_{eq}T_{eq}^4$ with $g_{eq}=3.36$ (see \cite{rg}), we  get 
$T_{eq}=3.25\times 10^{-28} M_{pl}=7.81\times 10^{-10}$ GeV. Thus, one easily obtains:
\begin{eqnarray}\label{xeq}
 \varphi_{eq}=\varphi_{rh}+2\sqrt{\frac{2}{3}}M_{pl}\left(1-\sqrt{\frac{2H_{eq}}{3H_{rh}}}\right)
 =\varphi_{rh}+2\sqrt{\frac{2}{3}}M_{pl}\left(1-\sqrt{\frac{2}{3}}\left( \frac{g_{eq}}{g_{rh}} \right)^{1/4}\frac{T_{eq}}{T_{rh}}\right) \nonumber \\
  \cong \varphi_{rh}+2\sqrt{\frac{2}{3}}M_{pl} 
  \cong 25.23 M_{pl}.
  \end{eqnarray}
 \begin{eqnarray}\label{yeq}
\dot{\varphi}_{eq}=\dot{\varphi}_{rh}\frac{t_{rh}}{t_{eq}}\sqrt{\frac{t_{rh}}{t_{eq}}}=\frac{4}{3}M_{pl}H_{eq}\sqrt{\frac{H_{eq}}{H_{rh}}}
=\frac{4\pi}{9}\sqrt{\frac{g_{eq}}{5}}\left(\frac{g_{eq}}{g_{rh}} \right)^{1/4}\frac{T_{eq}^3}{T_{rh}}\cong  2.3\times 10^{-19} \mbox{ eV}^2.
\end{eqnarray} 

From these results, we can see that taking $\gamma=10$ and $M\ll M_{pl}$  (for example $M=10^{-8}M_{pl}$) to have a kination regime before the end of inflation we have
\begin{eqnarray}
V(\varphi_{eq})\ll \frac{\dot{\varphi}_{eq}^2}{2}\sim 10^{-38} \mbox{ eV}^4,
\end{eqnarray}
and since $\rho_{eq}\cong 8.8\times 10^{-1} \mbox{eV}^4\sim \mbox{eV}^4$, one gets that at the matter-radiation equality one has
$\Omega_{\varphi, eq}\sim 10^{-38}$, which is completely different form { $\Omega_{\varphi_{sc}, eq}=\frac{1}{25}$}, showing once again that the
scalar field is not in the basin of attraction of the scaling solution.

\subsection{Numerical simulations during radiation}

To show numerically that  at the beginning of radiation the scalar field is not in the basin of attraction of the scaling solution, 
first of all,  we calculate the value of the redshift at the beginning of the radiation epoch
\begin{eqnarray}
1+z_{rh}=\frac{a_0}{a_{rh}}=\frac{a_0}{a_{eq}}\frac{a_{eq}}{a_{rh}}=
%(1+z_{eq})\frac{a_{eq}}{a_{rh}}=
(1+z_{eq})\left(\frac{\rho_{r,rh}}{\rho_{r,eq}}  \right)^{1/4}= (1+z_{eq})\left(\frac{g_{rh}}{g_{eq}}  \right)^{1/4}\frac{T_{rh}}{T_{eq}},
\end{eqnarray}
where  we have used that $\rho_{r,eq}=\rho_{r,rh}\left(\frac{a_{rh}}{a_{eq}} \right)^4$.

 As we have already explained, we use instant preheating as a mechanism of reheating leading  to a 
 reheating temperature $T_{rh}=10^9$ GeV, and thus,  $z_{rh}=-1+1.02\times 10^{22}$.

Moreover, at the beginning of radiation, the energy density of the matter will be
\begin{eqnarray}
\rho_{m,rh}=\rho_{m,eq}\left( \frac{a_{eq}}{a_{rh}} \right)^3=\rho_{m,eq}\left(\frac{\rho_{r,rh}}{\rho_{r,eq}}  \right)^{3/4}
=\frac{\pi^2}{30}g_{eq}\left( \frac{g_{rh}}{g_{eq}} \right)^{3/4}T^3_{rh}T_{eq}\nonumber \\ \cong 1.15\times 10^{19} \mbox{ GeV}^4.
\end{eqnarray}
Then,  to obtain the dynamical equations after the beginning of the radiation we use 
$N\equiv -\ln(1+z)=\ln\left( \frac{a}{a_0}\right)$ as a ``time'' variable. Now, using  the variable $N$,  one can recast the  energy density of radiation and matter respectively as, 
\begin{eqnarray}
\rho_{r}(a)={\rho_{r,rh}}\left(\frac{a_{rh}}{a}  \right)^4\Longrightarrow \rho_{r}(N)= {\rho_{r,rh}}e^{4(N_{rh}-N)} ,
\end{eqnarray}
and
\begin{eqnarray}
\rho_{m}(a)={\rho_{m,rh}}\left(\frac{a_{rh}}{a}  \right)^3\Longrightarrow \rho_{m}(N)={\rho_{m,rh}}e^{3(N_{rh}-N)},
\end{eqnarray}
where  $N_{rh}$ is the value of the time $N$ at the beginning of radiation, and  as we have already obtained, $\rho_{m,rh}\cong 1.15\times 10^{19} \mbox{ GeV}^4$ and 
$\rho_{r,rh}\cong 3.5\times 10^{37} \mbox{ GeV}^4$.
\begin{figure}
\begin{center}
\includegraphics[scale=0.5]{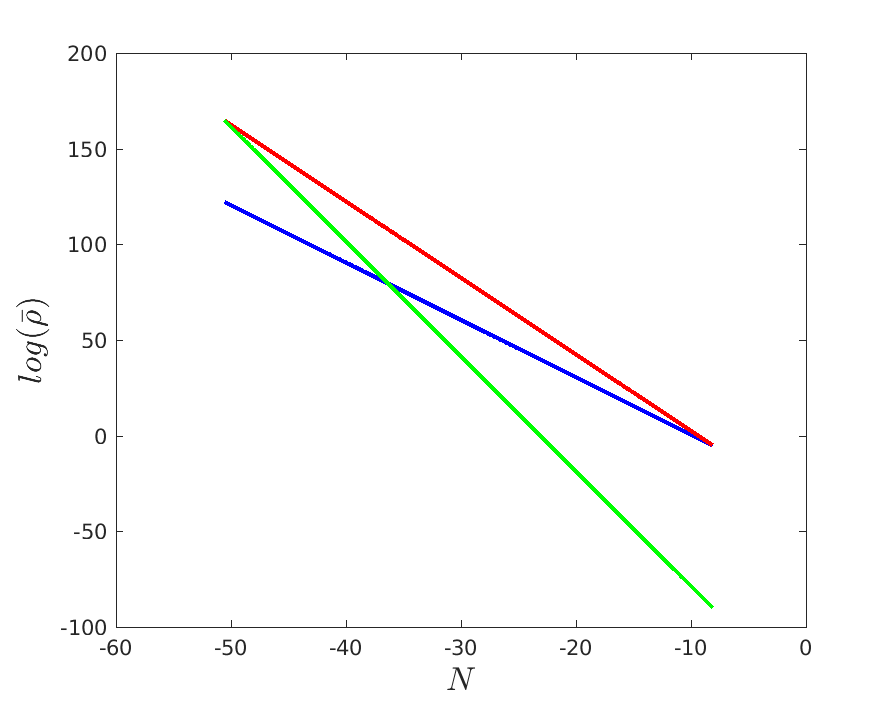}
\includegraphics[scale=0.54]{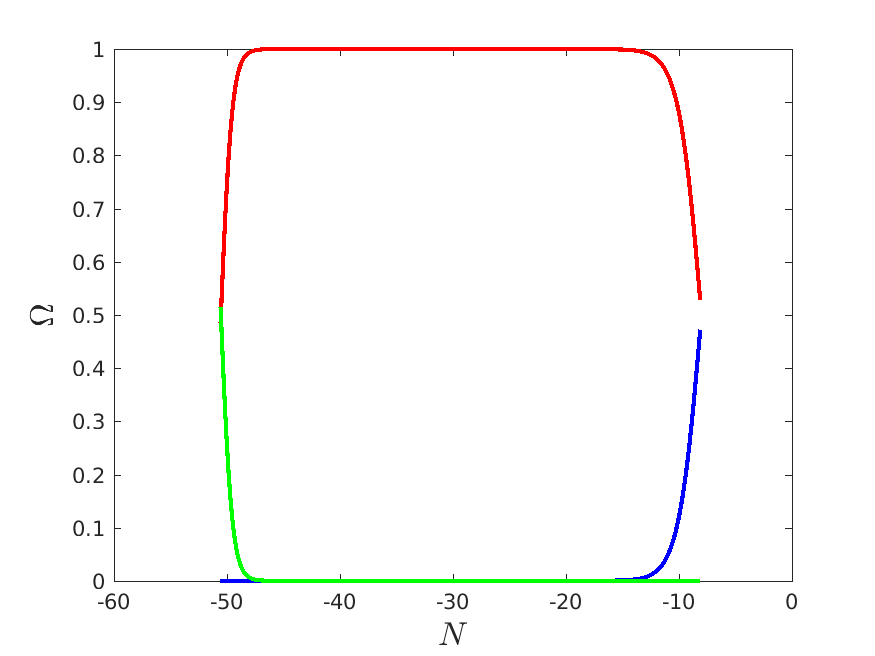}
\end{center}
\caption{{\em Left graph:} The reduced densities $\bar{\rho}_m$ (blue curve), $\bar{\rho}_r$ (red curve), and $\bar{\rho}_{\varphi}$ (green curve) in log scale, from reheating to matter-radiation equality have been shown. 
{\em Right graph:} The density parameters $\Omega_m, \Omega_r, \Omega_{\varphi}$ (using the same color code), from reheating to matter-radiation equality have been shown. Note that $\Omega_{\varphi,eq}<<\frac{1}{25}$, the value that it takes in case of  scaling solution. }
\label{fig:PV_rh_eq}
\end{figure}

\begin{figure}
\begin{center}
\includegraphics[scale=0.54]{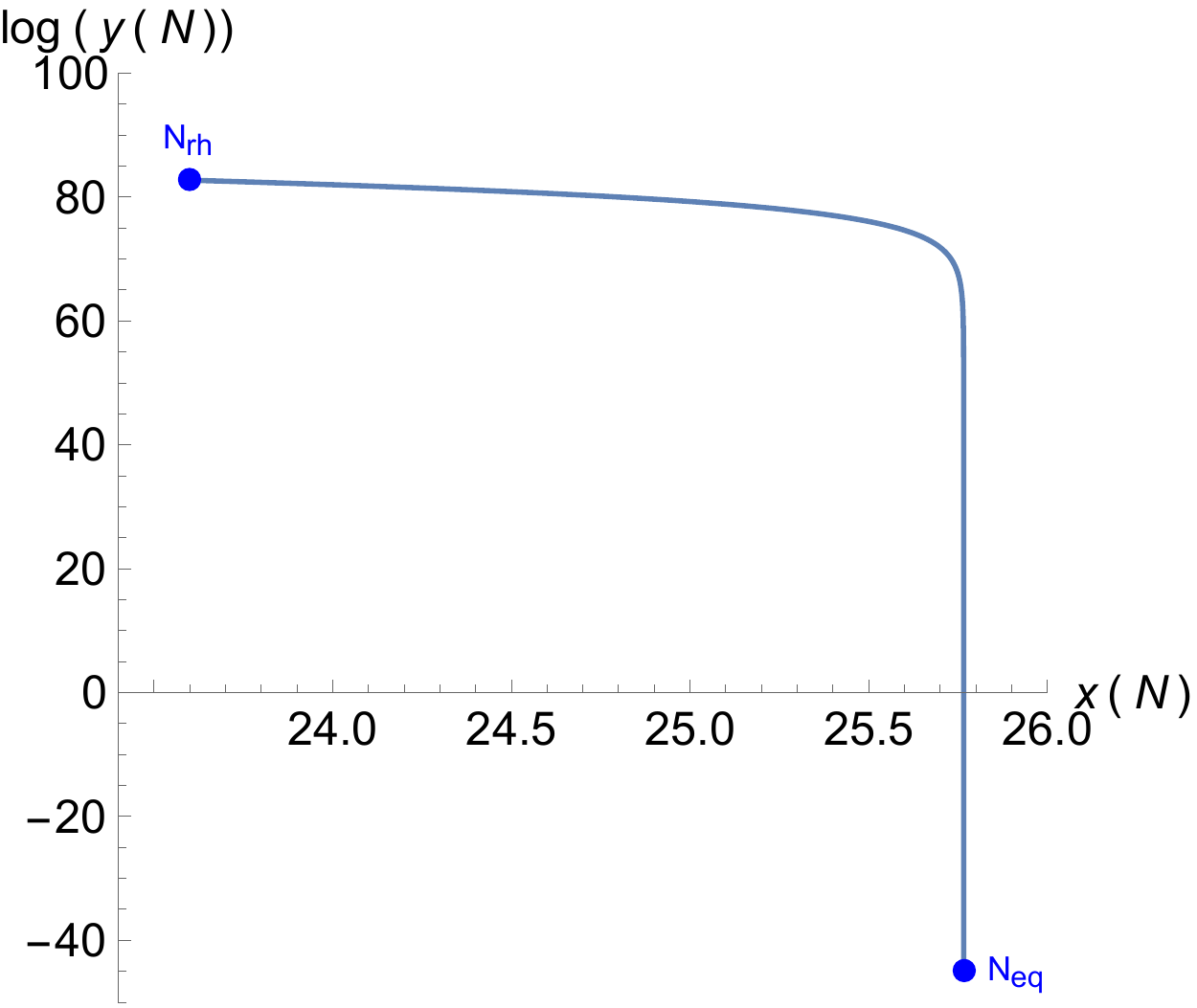}
\end{center}
\caption{Numerical integration of the dynamical system (\ref{system}) from the beginning of radiation to the matter-radiation equality, with initial condition at the beginning of reheating $x_{rh}=23.6$ and $y_{rh}=8.4\times 10^{35}$.}
\label{fig:NewPlot}
\end{figure}

Now, in order to obtain the dynamical system for this scalar field model, we 
introduce the following dimensionless variables
 \begin{eqnarray}
 x=\frac{\varphi}{M_{pl}}, \qquad y=\frac{\dot{\varphi}}{K M_{pl}},
 \end{eqnarray}
 where $K$ is a parameter that we have to choose accurately in order to facilitate the numerical calculations. Now, using the variable 
 $N = - \ln (1+z)$, defined above and also using the conservation equation $\ddot{\varphi}+3H\dot{\varphi}+V_{\varphi}=0$, one can construct the following  non-autonomous dynamical system:
 \begin{eqnarray}\label{system}
 \left\{ \begin{array}{ccc}
 x^\prime & =& \frac{y}{\bar H}~,\\
 y^\prime &=& -3y-\frac{\bar{V}_x}{ \bar{H}}~,\end{array}\right.
 \end{eqnarray}
 where the prime represents the derivative with respect to $N$, $\bar{H}=\frac{H}{K}$   and $\bar{V}=\frac{V}{K^2M_{pl}^2}$.  Moreover, 
 \begin{eqnarray}
 \bar{H}=\frac{1}{\sqrt{3}}\sqrt{ \frac{y^2}{2}+\bar{V}(x)+ \bar{\rho}_{r}(N)+\bar{\rho}_{m}(N) }~,
 \end{eqnarray}
where we have introduced the following dimensionless energy densities
 $\bar{\rho}_{r}=\frac{\rho_{r}}{K^2M_{pl}^2}$ and 
 $\bar{\rho}_{m}=\frac{\rho_{m}}{K^2M_{pl}^2}$.

When we integrate from the beginning of radiation up to the matter-radiation equality, i.e.,  from $N_{rh}\cong -50.57$ to $N_{eq}\cong -8.121$,
we choose $KM_{pl}= 10^{-17} \mbox{GeV}^2$, and thus, we get 
{\begin{eqnarray}
\bar{\rho}_{r}(N)= 3.5\times10^{71}e^{4(N_{rh}-N)}, \qquad \bar{\rho}_{m}(N)= 1.15 \times10^{53}e^{3(N_{rh}-N)}.
\end{eqnarray}}
We also choose $M=10^{-8} M_{pl}$, which returns
{\begin{eqnarray}
\bar{V}(x) \cong 2\times 10^{65} e^{-10 x}.
\end{eqnarray}}
The initial conditions for the field are $x_{rh}=23.6$ and $y_{rh}=8.4\times 10^{35}$.  The evolution of various cosmological parameters are graphically summarized in Fig. \ref{fig:PV_rh_eq}. Moreover, taking the same initial conditions as above, in Fig. \ref{fig:NewPlot}, we have shown the phase plane of the autonomous system (\ref{system}) from the beginning of the radiation era to the 
matter-radiation equality.   
A remarkable conclusion from our analsysi is that, at the matter-radiation equality time, i.e., $N_{eq}=-8.121$, the energy density of the field has a value 
$\Omega_{\varphi}(N_{eq})=5.84 \times 10^{-38}$  (which agrees with our theoretical result) which is completely different from the value of $1/25$, showing that it has  the scaling solution.

\section{A viable model}
\label{sec-IV}

As has been  shown in Ref. \cite{hap19} that for a viable model  with an exponential tail described in (\ref{SUSY}), the value of the model parameter $\gamma$ should belong to the interval $(0,\sqrt{2})$. It is important to realize that a value of $\gamma$ leading to a scaling solution does not serve for nothing because the real scalar field never enters the basin of attraction of the scaling solution. For this reason a potential with a double exponential character is also not necessary. What one needs is a tracker solution at late times, and as we can see in this case, the real scalar field is in the basin of attraction of this solution.

To obtain the dynamics we have to 
 integrate the dynamical system (\ref{system}) starting at $N_{eq}=-8.121$ with initial conditions (see eqs. (\ref{xeq}) and (\ref{yeq})): 
\begin{eqnarray}
x_{eq}=25.23 \qquad y_{eq}=2.3\times 10^{-19}\frac{\mbox{eV}^2}{KM_{pl}}
\end{eqnarray}
for the scalar field, and with 
\begin{eqnarray}
\bar{\rho}_{r}(N)=4.4\times 10^{-1}e^{4(N_{eq}-N)}\frac{\mbox{eV}^4}{K^2M_{pl}^2}~,\quad
\bar{\rho}_{m}(N)=4.4\times 10^{-1}e^{3(N_{eq}-N)}\frac{\mbox{eV}^4}{K^2M_{pl}^2}~,\end{eqnarray}
where in order to perform the numerical calculations we choose $KM_{pl}=10^{-4} \mbox{ eV}^2$.  Finally,  in order to see the dynamics of the  energy density of the scalar field, it is useful  to calculate the energy density of the tracker solution  (\ref{tracker}) as a function of the time $N$ which takes the form 
\begin{eqnarray}
\bar{\rho}_{tr}(N)=3\bar{H}_0^2e^{-\gamma^2 N},
\end{eqnarray}
and show numerically how both energy densities coincide at late times.

\begin{figure}
\begin{center}
\includegraphics[scale=0.45]{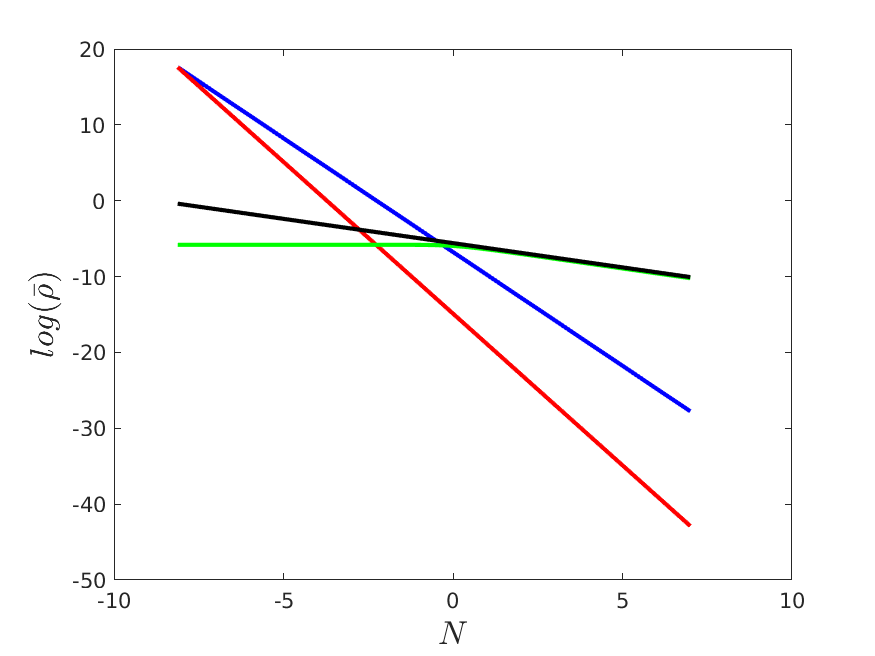}
\end{center}
\caption{Reduced densities $\bar{\rho}_m$ (blue curve), $\bar{\rho}_r$ (red curve), $\bar{\rho}_{\varphi}$ (green curve), and $\rho_{tr}$ (black curve) in log scale, from matter-radiation equality to the future have been shown. One can clearly see that after a certain value of $N$,  $\bar{\rho}_{\varphi}$ (green curve) coincides with the tracker solution $\bar{\rho}_{tr}$ (black curve).} 
\label{fig:rhob_eq_0}
\end{figure}
\begin{figure}
\begin{center}
\includegraphics[scale=0.45]{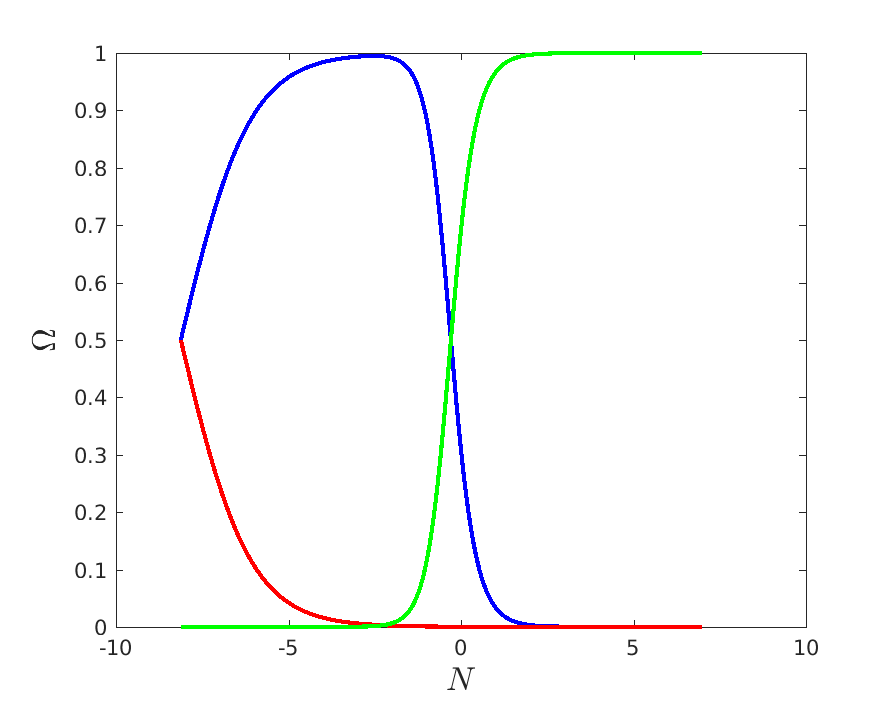}
\includegraphics[scale=0.45]{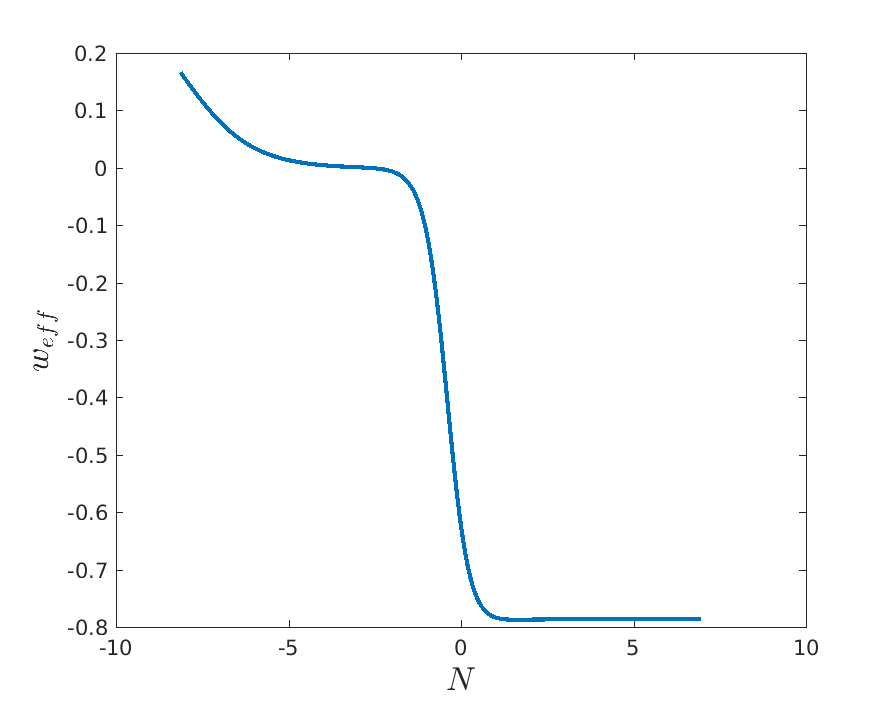}
\end{center}
\caption{{\em Left panel:} The density parameters $\Omega_m$ (blue curve), $\Omega_r$ (red curve), $\Omega_{\varphi}$ (green curve), from matter-radiation equality to the future have been depicted in the figure. {\em Right panel:} We show the qualitative evolution of the effective EoS $w_{eff}$, in the same time interval. One can see that for $N \rightarrow \infty$, (equivalently, $a \rightarrow \infty$), the effective EoS approaches towards a constant value equal to $\frac{\gamma^2}{3}-1\cong -0.786$. }
\label{fig:Omegaweff_eq_0}
\end{figure}

\subsection{Numerical simulations}

The integration of the system has been performed for $\gamma=0.8$. First of all, the value of the reduced mass
$\bar{M}$ that yields the observed value of $\bar{H}_0=3.53\times 10^{-2}$, has been found by application of the shooting method to be $\bar{M}=9.48 \times 10^{-4}$. Next, the system (\ref{system})
has been integrated from $N_{eq}=-8.121$ to future time $N=7$ with a RK78 numerical integrator. The upper limit has been chosen to verify the convergence of the solution to the tracker one.
The results are graphically illustrated in Fig. \ref{fig:rhob_eq_0} and Fig. \ref{fig:Omegaweff_eq_0}. In particular, in Fig. \ref{fig:rhob_eq_0}, we show the reduced densities of various cosmic fluids considering a wide range of the cosmic evolution, namely from the matter-radiation equality to the far future where we notice that the reduced energy density for the scalar field $\bar{\rho}_{\varphi}$ converges to the reduced energy density representing the tracker solution $\bar{\rho}_{tr}$. On the other hand, in Fig. \ref{fig:Omegaweff_eq_0} we plot the density parameters (left graph of Fig. \ref{fig:Omegaweff_eq_0}) and the effective equation of state (right graph of Fig. \ref{fig:Omegaweff_eq_0}). From the right graph of Fig. \ref{fig:Omegaweff_eq_0}, one can clearly visualize the overall evolution of the 
effective EoS that approaches towards a constant value in an 
asymptotic manner.

\section{Summary and Concluding remarks}
\label{sec-summary}

Scalar field models have uttermost importance in cosmic dynamics. Aside from their ability to explain the early acceleration (inflation) of the universe, they are also able to trace out the quintessence phase (late-time acceleration) of the universe. As already known, in standard inflation, after the decay of the inflation field and when the radiation phase just begins, if we consider a quintessence scalar field, then for an  exponential potential of the quintessence filed, the  scaling solutions exist. The scaling solutions have attractor behavior, however, to realize the present acceleration of the universe, the quintessence field has to leave the scaling behavior, and this can be done in several ways. For instance,  a double exponential potential
could play this role. On the other hand, the same features can be realized by introducing a non-minimal coupling between the quintessence 
field and massive neutrinos.

While on the other hand, the quintessential inflationary models are different from the usual scalar field models and hence they are worth in this context. In the present work we show that in the beginning of the radiation era, the scalar field is not in the basin of attraction of the scaling solution. Moreover, the value of 
$\Omega_{\varphi}$  becomes very low and this satisfies the bounds of BBN and recombination. Therefore, in quintessential inflation models we do not need any mechanism to exit the scaling behavior in contrary to the usual quintessence scalar field models. 
To illustrate this we have considered a viable quintessential inflationary model and shown the existence of both scaling and tracker solutions. For a better understanding we have graphically presented the behaviour of the model (see Figs. \ref{fig:rhob_eq_0} 
and \ref{fig:Omegaweff_eq_0}). From Fig. \ref{fig:rhob_eq_0}, it is clearly seen that $\bar{\rho}_{\phi}$ (green curve) coincides with $\bar{\rho}_{tr}$ (black curve) after a certain time and exhibits the current cosmic acceleration which is depicted through the behaviour of the effective EoS displayed in the right graph of Fig. \ref{fig:Omegaweff_eq_0}. In summary we conclude that the quintessential inflationary potentials are really impressive based on their outcomes and demands further investigations in light of the Planck's final likelihood release \cite{planck18,Aghanim:2019ame}.

\vspace{1cm}

{\it Acknowledgments:} 
The authors thank the referee for her/his critical reading of the manuscript and some important comments that helped us to improve the quality of the discussion.   
This investigation has been supported by MINECO (Spain) grants  MTM2014-52402-C3-1-P and MTM2017-84214-C2-1-P, and  in part by the Catalan Government 2017-SGR-247.
SP has been supported by the Mathematical Research Impact-Centric Support Scheme (MATRICS), File No. MTR/2018/000940, given by the Science and Engineering Research Board (SERB), Government of India.

%---------------------	
\end{document}